\documentstyle[12pt]{article}
\newlength{\extraspace}
\newlength{\extraspaces}
\newcommand{\be}{\begin{equation}
\addtolength{\abovedisplayskip}{\extraspaces}
\addtolength{\belowdisplayskip}{\extraspaces}
\addtolength{\abovedisplayshortskip}{\extraspace}
\addtolength{\belowdisplayshortskip}{\extraspace}}
\newcommand{\ee}{\end{equation}}

\newcommand{\ba}{\begin{eqnarray}
\addtolength{\abovedisplayskip}{\extraspaces}
\addtolength{\belowdisplayskip}{\extraspaces}
\addtolength{\abovedisplayshortskip}{\extraspace}
\addtolength{\belowdisplayshortskip}{\extraspace}}
\newcommand{\ea}{\end{eqnarray}}

\newcommand{\nonu}{\nonumber \\[.5mm]}
\newcommand{\A}{&\!\!\!}

\newcommand{\newsection}[1]{
\vspace{7mm} \pagebreak[3] \addtocounter{section}{1}
\setcounter{subsection}{0} \setcounter{footnote}{0}
\begin{center}
{\large {\bf \thesection. #1}}
\end{center}
\nopagebreak
\medskip
\nopagebreak \hspace{3mm}}

\begin{document}

\begin{center}
{{\bf Stability of Reissner Nordstr$\ddot{o}$m Black Hole}}
\end{center}
\centerline{ Gamal G.L. Nashed }

\bigskip

\centerline{{\it Mathematics Department, Faculty of Science, Ain
Shams University, Cairo, Egypt }}

\bigskip
 \centerline{ e-mail:nashed@asunet.shams.edu.eg}

\hspace{2cm}
\\
\\
\\
\\
\\
\\
\\
\\

The singularity of the solutions obtained  before in the
teleparallel  theory of gravitation is studied.  Also the
stability of these solutions is studied  using the equations of
geodesic deviation. The condition of stability is obtained. From
this condition the stability of  Schwarzschild solution can be
obtained.
\\\
\\
\\

\begin{center}
\newsection{\bf Introduction}
\end{center}

Static uncharged black holes in general relativity are described
by the well-known Schwarzschild solution. Singularity can be
happen when a gravitational collapse takes place and continue
until the surface of the star approaches the Schwarzschild radius,
i.e., $r=2m$ \cite{Di5}. Hawking and collaborators discovered that
the laws of thermodynamics have an exact analogues in the
properties of black holes \cite{Di5}$\sim$ \cite{Hw3}. As a black
hole emits particles, its mass and size steadily decrease. This
makes it easier to tunnel out and so the emission will continue at
an ever-increasing rate until eventually the black hole radiates
itself out of existence. In the long run, every black hole in the
universe will evaporates in this way.

The tetrad theory  of gravitation  based on the geometry of
absolute parallelism \cite{PP}$\sim$\cite{AGP1} can be considered
as the closest alternative to general relativity, and it has a
number of attractive features both from the geometrical and
physical viewpoints. Absolute parallelism is naturally formulated
by gauging space-time translations and underlain by the
Weitzenb$\ddot{o}$ck geometry, which is characterized by the
metricity  condition and by the vanishing of the curvature tensor
(constructed from the connection of the Weitzenb$\ddot{o}$ck
geometry). Translations are closely related to the group of
general coordinate transformations which underlies general
relativity. Therefore, the energy-momentum tensor represents the
matter source in the field equation for the gravitational field
just like in general relativity.

The tetrad formulation of gravitation was considered by M\o ller
in connection with attempts to define the energy of gravitational
field \cite{Mo8,Mo2}. For a satisfactory description of the total
energy of an isolated system it is necessary that the
energy-density of the gravitational field is given in terms of
first- and/or second-order derivatives of the gravitational field
variables. It is well-known that there exists no covariant,
nontrivial expression constructed out of the metric tensor.
However, covariant expressions that contain a quadratic form of
first-order derivatives of the tetrad field are feasible. Thus it
is legitimate to conjecture that the difficulties regarding the
problem of defining the gravitational energy-momentum are related
to the geometrical description of the gravitational field rather
than are an intrinsic drawback of the theory \cite{Mj,MDTC}.

In an earlier paper \cite{Naqc}, the author used a spherically
symmetric tetrad constructed by Robertson \cite{Ro} to derive
three different spherically symmetric space-times for the coupled
gravitational and electromagnetic fields with charged source in
the tetrad  theory of gravitation. One of these, contains an
arbitrary function and generates the others. These space-times
give the Reissner Nordstr$\ddot{o}$m metric black hole.
Calculations of
   the energy associated with these black holes  using the
superpotential method given by M\o ller \cite{Mo8} have been done
 \cite{Naqc}. It has been shown that unless the time-space components of the
tetrad field go to zero faster than ${1/\sqrt{r}}$ at infinity,
one got different results for the energy.

It is the aim of the present paper to study the singularity of the
three black hole solutions obtained before \cite{Naqc} and then
derive the condition of stability using the geodesic deviation
\cite{WB}. This study is important to gain more investigation
about the solutions obtained before \cite{Naqc}. In \S 2, a brief
review of the three black holes are given. The singularity problem
of these black holes  is studied in \S 3. In \S 4,  the condition
of stability is given. Final section is devoted to main results.

\begin{center}
\newsection{\bf Spherically symmetric black hole solutions}
\end{center}

In a previous paper the author used the teleparallel space-time in
which the fundamental fields of gravitation are the parallel
vector fields ${b_k}^\mu$. In the Weitzenb{\rm $\ddot{o}$}ck
space-time the fundamental field variables describing gravity are
a quadruplet of parallel vector fields \cite{HS} ${b_i}^\mu$,
which we call the tetrad field in this paper, characterized by \be
D_{\nu} {b_i}^\mu=\partial_{\nu} {b_i}^\mu+{\Gamma^\mu}_{\lambda
\nu} {b_i}^\lambda=0, \ee where ${\Gamma^\mu}_{\lambda \nu}$
define the nonsymmetric affine connection coefficients. The metric
tensor $g_{\mu \nu}$ is given by \[g_{\mu \nu}= {b^i}_\mu b_{i
\nu},\] where summation convention is taken over $i$. Equation (1)
leads to the metricity condition and the identically vanishing
curvature tensor.

 The gravitational Lagrangian $L_G$ is an invariant constructed
 from $g_{\mu \nu}$ and the contorsion tensor
  $\gamma_{\mu \nu \rho}$ given by \be \gamma_{\mu \nu \rho} =
{b^i}_{\mu}b_{i \nu; \ \rho}=\displaystyle{1 \over 2}\left(T_{\nu
\mu \rho}+T_{\rho \mu \nu}-T_{\mu \nu \rho}\right) \qquad T_{\mu
\nu \rho}={b^i}_{\mu} b_{i \nu, \ \rho}-{b^i}_{\nu} b_{i \mu, \
\rho}, \ee where the semicolon denotes covariant  differentiation
with respect to Christoffel symbols  and comma is the ordinary
differentiation and $T_{\mu \nu \rho}$ is the torsion. It is of
interest to note that the concept is as old as the gravitation
theory of Einstein. The torsion notion of a variety, besides the
curvature was introduced by Cartan \cite{Ce} that also gave a
geometric interpretation for both tensors. In teleparallel
theories the gravitational interaction is described by a force
similar to the Lorentz force equation of electrodynamics, with
torsion playing the role of force \cite{MDTC}.

The most general gravitational Lagrangian density invariant under
parity operation is given by the form \cite{HN,HS,MWHL}
 \be
{\cal
 L}_G  =  \sqrt{-g} L_G = \sqrt{-g} \left( \alpha_1 \Phi^\mu \Phi_\mu
 + \alpha_2 \gamma^{\mu \nu
\rho} \gamma_{\mu \nu \rho}+ \alpha_3 \gamma^{\mu \nu \rho}
\gamma_{\rho \nu \mu} \right) \ee
 with $g = {\rm det}(g_{\mu
\nu})$ and $\Phi_\mu$ being the basic vector field defined by
 $\Phi_\mu = {\gamma^\rho}_{\mu \rho}$.  Here $\alpha_1,
\alpha_2,$ and $\alpha_3$ are constants determined
 such that the theory coincides with general relativity in the weak
 fields \cite{Mo8,HS}:
\be
 \alpha_1=-{1 \over \kappa}, \qquad \alpha_2={\lambda \over
\kappa}, \qquad \alpha_3={1 \over \kappa}(1-\lambda), \ee
 where
$\kappa$ is the Einstein constant and  $\lambda$ is a free
dimensionless parameter\footnote{Throughout this paper we use the
relativistic units, $c=G=1$ and
 $\kappa=8\pi$.}.

The electromagnetic Lagrangian  density ${\it L_{e.m.}}$ is
\cite{KT}  \be {\it L_{e.m.}}=-\displaystyle{1 \over 4} g^{\mu
\rho} g^{\nu \sigma} F_{\mu \nu} F_{\rho \sigma}, \ee with $F_{\mu
\nu}$ being given by\footnote{Heaviside-Lorentz rationalized units
will be used throughout this paper} $F_{\mu \nu}=
\partial_\mu A_\nu-\partial_\nu A_\mu$.

The gravitational and electromagnetic field equations for the
system described by ${\it L_G}+{\it L_{e.m.}}$ are the following:

 \be G_{\mu \nu} +H_{\mu \nu} =
-{\kappa} T_{\mu \nu}, \qquad  K_{\mu \nu}=0, \qquad
\partial_\nu \left( \sqrt{-g} F^{\mu \nu} \right)=0. \ee
with $G_{\mu \nu}$ being the Einstein tensor of general relativity
defined by \be G_{\mu \nu}=R_{\mu \nu}-{1 \over 2} g_{\mu \nu} R,
\ee   R$_{\mu \nu}(\{\})$
 is the Ricci tensor defined by
\[ R_{\mu \nu}(\{\})= \partial_\rho \left\{ ^\rho_{\mu \nu} \right\}
 -\partial_\nu \left\{ ^\rho_{\mu \rho} \right \}+
\left\{ ^\rho_{\lambda \rho} \right \} \left\{ ^\lambda_{\mu \nu}
\right\}- \left\{ ^\rho_{\lambda \nu} \right\}
\left\{^\lambda_{\mu \rho} \right\}\] and $R(\{\})$ is the Ricci
scalar \[ R(\{\})=g^{\mu \nu} R_{\mu \nu} \]
 $H_{\mu \nu}$ and $K_{\mu \nu}$ are defined by \be H_{\mu \nu}
= \lambda \left[ \gamma_{\rho \sigma \mu} {\gamma^{\rho
\sigma}}_\nu+\gamma_{\rho \sigma \mu} {\gamma_\nu}^{\rho
\sigma}+\gamma_{\rho \sigma \nu} {\gamma_\mu}^{\rho \sigma}+g_{\mu
\nu} \left( \gamma_{\rho \sigma \lambda} \gamma^{\lambda \sigma
\rho}-{1 \over 2} \gamma_{\rho \sigma \lambda} \gamma^{\rho \sigma
\lambda} \right) \right],
 \ee
and \be K_{\mu \nu} = \lambda \left[ \Phi_{\mu,\nu}-\Phi_{\nu,\mu}
-\Phi_\rho \left({\gamma^\rho}_{\mu \nu}-{\gamma^\rho}_{\nu \mu}
\right)+ {{\gamma_{\mu \nu}}^{\rho}}_{;\rho} \right], \ee and they
are symmetric and antisymmetric tensors, respectively. The
energy-momentum tensor $T^{\mu \nu}$ is given by  \be T^{\mu
\nu}=-g_{\rho \sigma}F^{\mu \rho}F^{\nu \sigma}+\displaystyle{1
\over 4} g^{\mu \nu} F^{\rho \sigma} F_{\rho \sigma} \ee

It can be shown \cite{HS} that in spherically symmetric case the
antisymmetric part of the field equations (Eq.(6)) implies that
the axial-vector part of the torsion tensor, $a_\mu =
(1/3)\epsilon_{\mu\nu\rho\sigma}\gamma^{\nu\rho\sigma}$, should be
vanishing.  Then  $H_{\mu\nu}$ in Eq. (8) vanishes, and the field
equations (Eq.(6))  reduce to the coupled teleparallel equivalent
of Einstein-Maxwell equations. Equations (6) then determines the
tetrad field only up to local Lorentz transformations
\[
{b^k}_\mu \to {\Lambda(x)^k}_{\ell}\, {b^{\ell}}_\mu\,,
\]
 which retain the condition $a_\mu =0$.  Hereafter we shall refer to this
 property of the field equations as {\it restricted local Lorentz
 invariance}.

The structure of the Weintzenb${\ddot o}$ck spaces with spherical
symmetry and  three unknown functions of radial coordinate was
given by Robertson \cite{Ro}  in the form
 \be
\left({b_i}^\mu \right)= \left( \matrix{ iA & iDr & 0 & 0
\vspace{3mm} \cr 0 & B \sin\theta \cos\phi & \displaystyle{B \over
r}\cos\theta \cos\phi
 & -\displaystyle{B \sin\phi \over r \sin\theta} \vspace{3mm} \cr
0 & B \sin\theta \sin\phi & \displaystyle{B \over r}\cos\theta
\sin\phi
 & \displaystyle{B \cos\phi \over r \sin\theta} \vspace{3mm} \cr
0 & B \cos\theta & -\displaystyle{B \over r}\sin\theta  & 0 \cr }
\right), \ee where the vector ${b_0}^\mu$ has taken to be
imaginary in order to preserve the Lorentz signature for the
metric, i.e, the functions $A$ and $D$ have to be taken as
imaginary. Applying (11) to the field equations (Eq.(6)) the
author got \cite{Naqc} a set of non linear partial differential
equations. The solution of these equations has the form \cite{Naqc}:\\
\underline{First Solution}  \ba \A \A If \qquad  {A(R)} =
\displaystyle{1 \over \sqrt{1-\displaystyle{2m \over
R}+\displaystyle{q^2 \over R^2}}}, \qquad \qquad
 {B(R)} = \sqrt{1-\displaystyle{2m \over
R}+\displaystyle{q^2 \over R^2}} ,  \nonu
\A \A and \qquad { D(R)}=0, \qquad where \qquad R=\frac{r}{B} .\ea
\underline{Second Solution} \be If \qquad A=1, \qquad B=1 \qquad
and \qquad D(r)=\displaystyle{ \sqrt{2mr-q^2} \over r^2}.\ee
\underline{Third Solution} \be If \qquad {A(R)}= \displaystyle{1
\over \left(1-R  B'\right)}, \qquad and \qquad {D}(R)
=\displaystyle{1 \over 1-R B'} \sqrt{\displaystyle{2m \over
R^3}+{q^2 \over R^4 }+ \displaystyle{ B' \over R} \left(R B' -2
\right)}. \ee It is clear from (14) that the third solution
depends on the arbitrary function ${ B}$, i.e., we can generate
the pervious solutions of (12) and (13) by choosing the arbitrary
function $B$ to have the form \be {B(R)} = 1, \qquad and \qquad
{B(R)}={\int{{1 \over R} \left(1-\sqrt{1-\displaystyle{2m \over
R}-\displaystyle{q^2 \over R^2}} \right)}}dR. \ee The associated
metric of the three solutions $(12)\sim(14)$  is found to be the
same and have the form \be ds^2=-\eta(r)dT^2+\displaystyle{dr
\over \eta(r)}+r^2d\Omega^2,\qquad  with \qquad
\eta(r)=\left[1-{2m \over r}+{q^2 \over r^2}\right], \ee which is
the  static Reissner Nordstr$\ddot{o}$m black hole \cite{DGA, NB}.
The form of the vector potential $A_\mu$,  the antisymmetric
electromagnetic
 tensor field  $F_{\mu \nu}$ and the energy-momentum tensor are given
 by \cite{Naqc}
  \be A_t(r)=-\displaystyle{q \over 2\sqrt{\pi} r}, \qquad
  F_{r t}=-\displaystyle{q \over 2\sqrt{\pi} r^2},\qquad
  {T_0}^0={T_1}^1=-{T_2}^2=-{T_3}^3=\displaystyle{q^2 \over 8 \pi
  r^4}.
  \ee
It is of interest to note that the two tensors $H_{\mu \nu}$ and
$K_{\mu \nu}$ are vanishing identically for the three solutions
given by Eq. (12), (13) and (14). It is proved that these tensors
are vanishing identically for any spherically symmetric solutions
\cite{HSN,Ngr}.
\newsection{Singularities}
In teleparallel theories we mean by singularity of space-time
\cite{KT} the singularity of the scalar concomitants of the
torsion and curvature tensors.

Using the definitions  of the  Riemann-Christoffel curvature
tensor, Ricci tensor, Ricci scalar, torsion tensor, basic vector,
traceless part and the axial vector part \cite{Nc2} we obtain for
the first solution of (12) \ba
 R^{\mu \nu \lambda \sigma}R_{\mu \nu \lambda
\sigma} \A = \A \displaystyle{ 8  \over R^8} \left[ 7q^4-12 R M
q^2+6 M^2 R^2 \right], \qquad  R^{\mu \nu}R_{\mu \nu} =
\displaystyle{ 4 q^4 \over R^8}, \qquad R = 0,\nonu
T^{\mu \nu \lambda}T_{\mu \nu \lambda} \A=\A \displaystyle{-2
\over R^4(R^2-2MR+q^2)} \Biggl[
4R^4-12MR^3+6R^2q^2-4R^3\sqrt{R^2-2MR+q^2}-10RMq^2 \nonu
 \A \A +8R^2\sqrt{R^2-2MR+q^2}M-4R\sqrt{R^2-2MR+q^2}q^2+9M^2R^2+3q^4
\Biggr ], \nonu
\Phi^\mu \Phi_\mu \A=\A \displaystyle{-1 \over R^4(R^2-2MR+q^2)}
\Biggl[ -2R^3+4MR^2-2Rq^2+2\sqrt{R^2-2MR+q^2}R^2\nonu
\A \A -3\sqrt{R^2-2MR+q^2}MR+\sqrt{R^2-2MR+q^2}q^2 \Biggr]^2,\nonu
t^{\mu \nu \lambda}t_{\mu \nu \lambda} \A=\A \displaystyle{-1
\over R^4(R^2-2MR+q^2)}
\Biggl[R^3-2MR^2+Rq^2-\sqrt{R^2-2MR+q^2}R^2 \nonu
 \A \A
 +3\sqrt{R^2-2MR+q^2}MR-2\sqrt{R^2-2MR+q^2}q^2\Biggr]^2,\qquad
a^\mu a_\mu = 0.
 \ea
The scalars of the Riemann-curvature tensor, Ricci tensor and
Ricci scalar of  the second solution (13) are the same as given by
(18). This is a logic results since both solutions reproduce the
same metric tensor and these scalars mainly depend on the metric
tensor. The scalars of torsion tensor, basic vector, traceless
part and the axial vector part of the  space-time given by
solution (13) are given by

\ba T^{\mu \nu \lambda}T_{\mu \nu \lambda} \A=\A\displaystyle{-2
\over (2Mr-q^2)r^4} \left[(3q^4-10 q^2Mr+9M^2r^2) \right], \qquad
\Phi^\mu \Phi_\mu = \displaystyle{-1 \over (2Mr-q^2)r^4} \left[
(3Mr-q^2) \right], \nonu
t^{\mu \nu \lambda}t_{\mu \nu \lambda} \A=\A \displaystyle{-1
\over (2Mr-q^2)r^4} \left[ (3Mr-2q^2) \right].
 \ea

It is clear  from (18) and (19) that the scalars of the torsion,
basic vector and the traceless part of the first two  solutions
given by Eqs. (12) and (13) are quite different in spite that they
gave the same associated metric (16)! The singularity of the
scalars of Riemann-curvature tensor, Ricci tensor and Ricci scalar
is given at $R\rightarrow 0$ and this is well know from general
relativity as we can see from Eq. (18) \cite{Di5}. As the
singularities of the scalars of the torsion, basic vector and the
traceless part of the first solution  (12) are $r \rightarrow 0$
and $R^2-2Mr+q^2\rightarrow 0$ the second singularity may has the
form $R \rightarrow M \pm \sqrt{M^2-q^2}$ which is the horizons of
the static Reissner Nordstr$\ddot{o}$m black hole  \cite{KE}.

The singularities of the second solution (19) are given by $r
\rightarrow 0$ and $q^2/2M$. Now we have two solutions reproduce
the same metric but the singularity of their space-times are not
coincide. This is expected of course due to the following facts:\\
i) The energy content of these space-times are different
\cite{Naqc}.\\  ii) The time-space components of the tetrad fields
${b_0}^\alpha, \ {b_\alpha}^0$ go to zero as $ \displaystyle{1
\over \sqrt{r}}$ at infinity \cite{HSN,Ngr}.
\vspace{.3cm}\\\hspace*{-.3 cm}  iii) Also another
interpretations, which may be taken into account as is clear from
Eqs. (18) and (19) is that the torsion tensor of these solutions
is different. As we discussed in the introduction that the torsion
plays the role of the force, therefore, we may interpret the
different results of the two torsions given by Eq. (18) and (19)
due to the fact that the forces of the two solutions are
different.

 \newsection{The Stability condition}
 In the background of gravitational field the trajectories are
 represented by the geodesic equation
 \be
 {d^2 x^\lambda \over ds^2}+ \left\{^\lambda_{ \mu \nu} \right \}
 {d x^\mu \over ds}{d x^\nu \over ds}=0,
 \ee
 where $\displaystyle{d x^\mu \over ds}$ is the velocity four vector,
 s is a parameter varying along the geodesic. It is well know that the
 perturbation of the geodesic will leads to deviation \cite{Di5}
 \be
 {d^2 \zeta^\lambda \over ds^2}+ 2\left\{^\lambda_{ \mu \nu} \right \}
 {d x^\mu \over ds}{d \zeta^\nu \over ds}+
 \left\{^\lambda_{ \mu \nu} \right \}_{,\ \rho}
 {d x^\mu \over ds}{d x^\nu \over ds}\zeta^\rho=0,
 \ee
where $\zeta^\rho$ is the deviation 4-vector.

Using Eqs. (20) and  (21) in (16) we get for the geodesic
equations \be {d^2 t \over ds^2}=0, \qquad {1 \over 2}
\eta'(r)\left({d t \over ds}\right)^2-r\left({d \phi \over
ds}\right)^2=0, \qquad {d^2 \theta \over ds^2}=0,\qquad {d^2 \phi
\over ds^2}=0,\ee  and for the geodesic deviation \ba \A \A {d^2
\zeta^0 \over ds^2}+{\eta'(r) \over \eta(r)}{dt \over ds}{d
\zeta^1 \over ds}=0, \nonu
\A \A  {d^2 \zeta^1 \over ds^2}+\eta(r)\eta'(r) {dt \over ds}{d
\zeta^0 \over ds}-2r \eta(r) {d \phi \over ds}{d \zeta^3 \over
ds}\nonu
\A \A +\left[{1 \over 2}\left(\eta'^2(r)+\eta(r) \eta''(r)
\right)\left({dt \over ds}\right)^2-\left(\eta(r)+r\eta'(r)
\right) \left({d\phi \over ds}\right)^2 \right]\zeta^1=0,\nonu
\A \A  {d^2 \zeta^2 \over ds^2}+\left({d\phi \over ds}\right)^2
\zeta^2=0, \qquad \qquad  {d^2 \zeta^3 \over ds^2}+{2 \over
r}{d\phi \over ds} {d \zeta^1 \over ds}=0, \ea where $\eta(r)$ is
defined by (16), $\eta'(r)=\displaystyle{d\eta(r) \over dr}$ and
we have consider the circular orbit in the plane \be \theta={\pi
\over 2}, \qquad {d\theta \over ds}=0, \qquad  {d r \over ds}=0.
\ee Using (24) in (16) we get \be \eta(r)\left({dt \over
ds}\right)^2-r^2\left({d\phi \over ds}\right)^2=1, \ee from (25)
and (22) we obtain \be \left({d\phi \over ds}\right)^2={\eta'(r)
\over r(2\eta(r)-r\eta'(r))}, \qquad \left({dt \over
ds}\right)^2={2 \over 2\eta(r)-r\eta'(r)}. \ee

The variable $s$ in (23) can be eliminated and we can rewrite it
in the form \ba \A \A {d^2 \zeta^0 \over d\phi^2}+{\eta'(r) \over
\eta(r)}{dt \over d\phi}{d \zeta^1 \over d\phi}=0 \nonu
\A \A {d^2 \zeta^1 \over d\phi^2}+\eta(r)\eta'(r) {dt \over
d\phi}{d \zeta^0 \over d\phi}-2r \eta(r) {d \zeta^3 \over
d\phi}\nonu
\A \A +\left[{1 \over 2}\left(\eta'^2(r)+\eta(r) \eta''(r)
\right)\left({dt \over d\phi}\right)^2-\left(\eta(r)+r\eta'(r)
\right)  \right]\zeta^1=0, \nonu
\A \A {d^2 \zeta^2 \over d\phi^2}+\zeta^2=0, \qquad  {d^2 \zeta^3
\over d\phi^2}+{2 \over r} {d \zeta^1 \over d\phi}=0. \ea It is
clear from the third equations of (27) that it represent a simple
harmonic motion, this means that the motion in the plan
$\theta=\pi/2$ is stable.

Assuming now the solution of the remaining equations given by \be
\zeta^0 = A_1 e^{i \omega \phi}, \qquad \zeta^1= A_2e^{i \omega
\phi}, \qquad and \qquad \zeta^3 = A_3 e^{i \omega \phi},\ee where
$A_1, A_2$ and $A_3$ are constants to be determined. Inserting
(28) in (27) we get \be  \displaystyle{r^3m-6m^2r^2+9mrq^2-4q^4
\over r^2(mr-q^2) }>0, \ee which is the condition of the stability
for a static spherically symmetric  Reissner Nordstr$\ddot{o}$m
solution. Condition (29) can be rewritten as \be
r-\displaystyle{q^2 \over m}>0 \qquad \qquad and \qquad \qquad
r-6m>0.\ee

\newsection{Main results}
The main results can be summarized  as follows \vspace{.3cm} \\ 1)
The singularity problem of the first two solutions (12) and (13)
obtained before \cite{Naqc} has been studied. The scalars of the
torsion tensor, basic vector and the traceless part of these
solutions  are quite different as we can see from Eq. (18) and
(19). The scalars  have a common singularity if $r\rightarrow 0$.
Furthermore, the first solution  has another singularity if
\[ r^2-2rm+q^2 \rightarrow 0,\] while the second solution  has
another singularity if  \[r -\displaystyle{q^2 \over
2M}\rightarrow 0.
\] This explains that the structure of the two  solutions (12) and (13) are
 quite different in spite that they reproduce the same
metric space-time.

2) The stability condition for the metric of Reissner
Nordstr$\ddot{o}$m black hole  Eq. (16) is
derived and is given by Eq. (29). From this condition we can see that.\vspace{.3cm}\\
i) If $r \rightarrow 0$ the value of (29) is finite.\vspace{.3cm}\\
ii) If r becomes large  then  Eq. (30) takes the value $r>6m$ and
$r>q^2/m$ which is the condition of  stability for Reissner
Nordstr$\ddot{o}$m black hole.\vspace{.3cm}\\ iii) When $q=0$ and
if r becomes large  then Eq. (32) takes the value $r>6m$ which the
is condition of  stability for Schwarzschild black hole
\cite{Nc2}. The analysis given here for the derivation of the
stability condition is a straightforward and so simple than that
used in the literature \cite{Vs,Wr}

\end{document}